# Multiple vibro-polaritons formation from a thin polyethylene film embedded in a resonant mid-infrared cavity

Mario Malerba[1], Mathieu Jeannin[1], Paul Goulain[1], Adel Bousseksou[1], Raffaele Colombelli[1], Jean-Michel Manceau [1,*]

1. Centre de Nanosciences et Nanotechnologies, Université Paris-Saclay, CNRS, UMR 9001, 91120 Palaiseau, France

We experimentally resolve the dispersion of multiple vibro-polariton modes issued from the strong coupling of different vibrational bands of the methylene group ($CH_2$) in a 2.56 µm thick polyethylene film with the confined modes of a mid-infrared Fabry-Perot micro-cavity. We measure a Rabi frequency of 111 $cm^{-1}$ for the stretching doublet around 2950 $cm^{-1}$ and a Rabi frequency of 29 $cm^{-1}$ for the scissoring doublet around 1460 $cm^{-1}$. This simple experimental approach offers the possibility to accurately fit the measured molecular film dielectric function. We show that the polariton dispersion and Rabi splitting can be precisely predicted from numerical simulations, offering a valuable tool for the design of strongly coupled system and the development of novel molecular films with crystalline organization.

1. **Introduction**

The strong light matter coupling of molecular systems with resonant photonic micro-cavities has attracted a lot of attention over the last decade due to potential applications in physics and chemistry [1,2]. Such regime of interaction occurs when the coupling strength between a matter excitation and a confined photonic mode is higher than their respective damping rates. Hybrid light-matter states appear that are commonly named polaritons. At NIR and visible wavelengths, the excitonic transition of organic materials are placed in cavity to enter the strong coupling regime and have been successful in demonstrating room temperature Bose-Einstein condensate and conductivity enhancement [3,4]. In stark contrast to the latter, when organic materials are used at mid-infrared wavelengths, cavities are made resonant with the vibrational bands of molecules. In fact, the recent years have witnessed a surge of activity on coupling the vibrational modes of organic systems with micro-cavities, to enter the so-called vibrational strong-coupling (VSC) regime. Such interaction usually occurs at wavelengths spanning from mid-IR ($\lambda$-3-30 µm) up to far-IR ($\lambda$ = 30-100 µm) [5–7], and has already been demonstrated with various solid-phase molecular systems, but also under liquid phase within a microfluidic cavity [8]. This regime offers important perspective from a chemistry point of view [7], with for instance the recent demonstration of gaining control over a reaction selectivity without illumination [10].

Ultrafast pump-probe spectroscopy, that aims at exploring the intermolecular vibrational energy transfer and relaxation within these dressed vibrational states, has also been explored [11,12]. Along with the novel possibilities brought on the control of molecular properties, these systems also appear as potential candidates for photonic devices, with the exploration of parametric processes under resonant optical pumping scheme. The perspective is to demonstrate Bose-Einstein condensation or novel types of optical parametric oscillators where the Raman active mode is dressed with a secondary micro-cavity mode, as proposed for instance in [2,13,14].

In this respect, gaining a precise control over the design and the fabrication of the cavity, as well as planarity of the molecular film placed within the cavity is of great importance. So far, little effort has been devoted to the optimization, design and precise measurement of the strongly coupled system. Few attempts to improve the lifetime of the photonic system using Bragg mirrors have been reported [15,16] along with approaches where the resonant EM mode is highly confined using nanoresonators supporting localized surface plasmon or phonon-polariton resonances [17–20].

In the present article, we report on the experimental demonstration of strong light-matter coupling of multiple vibrational absorption bands from a polyethylene (PE) film embedded within a Fabry-Perot (FP) micro-cavity. Polyethylene has a long history as a Raman active material and present sharp vibrational bands at mid-IR wavelengths [21–24]. Remarkably, the multiple polaritons formation occurs at the scissoring and stretching bands of the methylene group ($CH_2$). The FP cavity is formed by a fully reflective, thick bottom mirror and a semi-transparent top mirror, to operate only in reflection geometry. Spectroscopic measurements on the bare film deposited on the bottom mirror allow a precise quantitative extraction of the complex dielectric function of the PE film via Kramers-Kronig relation. The latter is then fitted with a multiple Lorentz oscillator model that is used within numerical simulations to accurately reproduce the apparition of the new eigenstates along with their spectral dispersion. We further use an analytical expression to estimate the minimum Rabi-splitting. Using the fitted permittivity data, we show that we can accurately predict the behaviour of a randomly organized molecular film in cavity. Our single port experiment appears as a valuable tool for the precise design and characterisation of optically pumped vibro-polaritonic systems and particularly relevant for the exploration of different molecular film organizations.

2. **Sample fabrication**

As active vibrational material, we have chosen one of the most common commercial thermoplastic, namely the mid-density polyethylene (MDPE, Sigma-Aldrich), a polymer made of long hydrocarbon chains of ethylene monomer ($C_2H_4$). The molecular film is obtained from a classic spin coating approach. Molecules under solid form are dissolved in decahydronaphtalene, a solvent than can be heated up to 190 C°. The melting temperature of polyethylene occurs at around 140 C°. The solution is spin-coated on a glass substrate metallized with a 200 nm thick gold film which constitutes the bottom mirror of the FP cavity. The glass substrate is kept at high temperature during the coating process to avoid precipitation of condensed solid flakes of MDPE during the spinning and maintain the desired viscosity ensuring reproducibility in the film thickness. After spin-coating, the sample is annealed at 150 C° in order to fully evaporate all the solvent and is then abruptly tempered into cold water in order to keep the amorphous character within the film organization. The sample is at last diced in two pieces of equal size. A 25 nm thick, semi-transparent gold mirror is finally deposited on one of them while the other is left uncoated to measure the absorbance of the PE film in a reflection geometry (Fig. 1a). Using a reflectometer, we determined a film thickness of 2.56 μm.

3. **Experimental measurement of the polaritonic dispersion**

We spectrally characterised our samples using a Fourier transform infrared spectrometer equipped with a reflection angle motorized unit that allows the precise investigation of the dispersion as a function of the impinging light angle. We used the SiC globar source in conjunction with a KBr (Potassium Bromide) beam splitter and a room temperature DTGS (Deuterated Triglycine sulfate) detector. We can control the polarization of the impinging light using a KRS-5 (Thallium Halogenide) grid polarizer as depicted in figure 1. We first probe the bare MDPE film at an incidence angle of 45 degree. As presented in figure 2 (left panel), the methylene group ($CH_2$), present in the MDPE polymeric film, exhibits two distinctive absorption bands that are evenly spaced in frequency, the scissoring modes at ~1460 $cm^{-1}$ and the stretching modes at ~2920 $cm^{-1}$, as already assigned in a vast body of literature [21–24]. They are doublets oscillating at a symmetric and anti-symmetric resonance. Note that a third doublet is present at a frequency around 730 $cm^{-1}$ which is known in the literature as the rocking mode (not shown here for clarity). We then probed the reflectivity $R(\omega,\theta)$ of the FP cavity over a large spectral bandwidth (600 to 7000 $cm^{-1}$) and a wide angular range (13° < $\theta$ < 73°). The impinging radiation is TM (Transverse Magnetic) polarized, ensuring a stronger contrast of the photonic mode. Measurement with TE (Transverse Electric) and unpolarized impinging radiation are provided in supplementary information (Figure S1) for the sake of completeness. The absolute reflectivity is obtained by division of the sample spectrum with a reference one taken from a planar gold surface as shown in [25]. The experimental dispersion of the polaritonic modes is presented in Figure 2 (right panel). The eigenstates originating from the strong coupling regime are clearly resolved and exhibit the characteristic anti-crossing signature of the polaritonic modes. The scissoring and the stretching band doublets are coupled to the 1st and 2nd order FP modes, showing a simultaneous coupling of different vibrational bands within the ethylene monomer. While three distinct polaritonic branches are observed in the case of the stretching band, we were able to resolve only two branches in the case of the scissoring mode even with an increased resolution of 2cm$^{-1}$, due to the very weak contrast of the central branch.

4. **Fitting model for the single port system**

Our single-port experimental geometry enables an accurate estimation of the dielectric function of the bare PE film ($n$, $k$). From the absorbance measurement on the bare film on Au, one can use Beer-Lambert's law to deduce the imaginary part of the refractive index. Using Kramers-Kronig relation, the real part of the refractive index function can be calculated over our frequency range of interest as described in [26,27] and the complex permittivity function can be retrieved. We then fit these data with a multiple Lorentz permittivity model using the following expression:

$$\varepsilon(\omega) = \varepsilon_b - \sum_{k=1}^{n} \frac{2 a_k \omega_k}{\omega^2 - \omega_k^2 + i\omega\gamma_k}$$

We use the formalism developed in [28], where $a_k$, $\omega_k$ and $\gamma_k$ are the amplitude, the frequency and the decay rate of each resonance. The background permittivity, $\varepsilon_b$, is equal to 2,2. Figure 3, shows the excellent agreement in between the extracted permittivity and the Lorentz model for both vibrational set of doublets. We provide the fitting values within the table presented in supplementary information. In fact, the stretching doublet is better fitted with four distinct Lorentzian curves. Two modes (stretching 2 and 4 within the table) with lower amplitude and broader linewidth allow to better grasp the overall line shape of this vibrational doublet.

Interestingly, this fitted permittivity can be subsequently plugged into numerical simulations to depict the organic polaritonic system with high accuracy. Figure 4a shows the numerical simulations obtained from a rigorous coupled wave analysis code (RCWA) [29,30], where the fitted permittivity has been used to account for the macroscopic randomly organized molecular film embedded within a metal-metal FP micro-cavity. The gold permittivity is taken from [31]. The experimental reflectivity minima are superimposed (white dots) on the reflectance colour-plot and their spectral positions are in excellent agreement with the numerical simulation. The linewidth of the polaritonic branches is significantly smaller with respect to the experiment, while the trend of the resonant mode contrast is similar. Two factors could explain this difference: (i) the angular broadening from the measurement system in contrast with the plane-wave used in numerical simulation (ii) spatial variations in PE film thickness, which strongly broaden the system under measurement. As a matter of fact, we have been able to probe the film thickness at different locations on the sample using a microscope FTIR (Figure S2) with a 150 μm sided square spot area. Using the analytical expression of a FP resonant mode $\nu_k = kc/2nL$, where L is the thickness of the film and n the refractive index, we have estimated using the 3$^{rd}$ order FP mode, a disparity of about 160 nm in thickness over a 1mm$^2$ sample area. This observation contrasts with AFM (atomic force microscope) measurements of the local roughness that have shown a film roughness of about 10 nm over tens of microns length. Such long-range modulation of the film thickness could be the direct result of the spin-coating technique that often generate ripples. In turn, gaining a higher degree of control over the film thickness would be beneficial for photonic applications where the dressed state lifetime is involved and technique such as dip coating could be employed [32].

Finally, we can use the formalism developed in [28], which links quantum formulation and classical calculation, to plug the fitted values within the following analytical expression that gives the minimum Rabi splitting frequency:

$$\Omega_S = \sum_{k=1}^{n} \sqrt{\frac{a_k \omega_k}{\epsilon_b}} f_w$$

Where $f_w$ is an overlap factor between the confined EM mode and the active molecular film. In the present case of a FP cavity, this value is equal to 1. Starting with the stretching modes, we have experimentally measured a Rabi splitting of 212 cm$^{-1}$ between the lower and upper polaritonic states while the above calculation gives a value of 215 cm$^{-1}$. For the scissoring modes, we extracted an experimental Rabi splitting of 46 cm$^{-1}$ while the above expression gives 38 cm$^{-1}$. This further confirms that the single port experimental approach we have presented is a powerful tool in view of designing and optimizing molecular films. With accurate control over the bare molecular film thickness, one can precisely predict the expected Rabi splitting frequency – and the full dispersion - once the film is placed in the micro-cavity. Therefore any attempt to optimize the polymeric orientation to obtain - for instance - a crystal like orientation with a higher number of aligned dipoles could be gauged rapidly and precisely with a simple reflectance measurement [21,22].

5. Conclusion

In conclusion, we have shown multiple and simultaneous polaritons formation from different vibrational modes of the same ethylene monomer in conjunction with the 1$^{st}$ and 2$^{nd}$ order EM modes of a metal-metal FP micro-cavity. A simple reflectance measurement allows to accurately extract the dielectric function of the bare molecular film which can be fitted with a simple Lorentz oscillator permittivity model. The Rabi splitting frequency can be easily predicted and the polaritons dispersion precisely calculated with numerical simulations. The formation of polaritonic modes involving different photonic and vibrational modes could be a useful resource for non-linear experiments exploiting e.g. vibrational intermode coupling or multiply-resonant features.

**Acknowledgments:** We warmly thank M. Manceau, A. Cattoni and J. Feist for useful discussions while initiating this research activity. We acknowledge financial support from the European Union FET-Open Grant MIRBOSE (737017) and from the French National Research Agency (Projects ANR-19-CE24-0003 "SOLID" and ANR-17-CE24-0016 "IRENA"). This work was partially supported by the French RENATECH network.

**\*Corresponding author**: jean-michel.manceau@c2n.upsaclay.fr

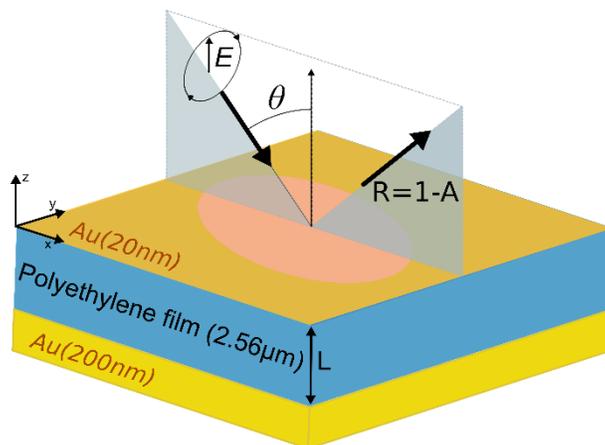

Figure 1: Schematic representation of the experimental single approach. The ground mirror is thick enough (200nm) so the system can only be probed in reflection using a motorized angle unit inserted within an FTIR spectrometer. When embedded in cavity, the PE film is covered with a semi-transparent gold mirror of 20 nm thickness.

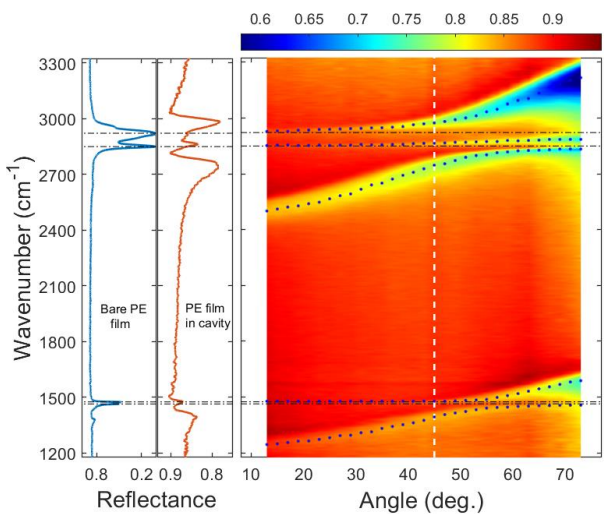

Figure 2: (Left panel) *In blue,* reflectance of the bare PE film spin-coated on the gold surface taken at an angle of 45 degrees. Both scissoring and stretching doublets of the ethylene monomer ($CH_2$) can be resolved with an experimental resolution of 2cm$^{-1}$. *In orange*, reflectance of the same film embedded in cavity, taken at 45 degrees incidence. (Right panel) Experimental angle-resolved reflectance of the 2.56μm PE film embedded in the double metal FP micro-cavity. Both scissoring and stretching vibrational modes are coupled to the $n_1$ and $n_2$ FP cavity modes. We can resolve the polaritonic branches dispersion. The blue dots correspond to the reflectance minima. The white dash line indicate the 45 degree angle at which the presented reflectance cuts on the left hand-side were done.

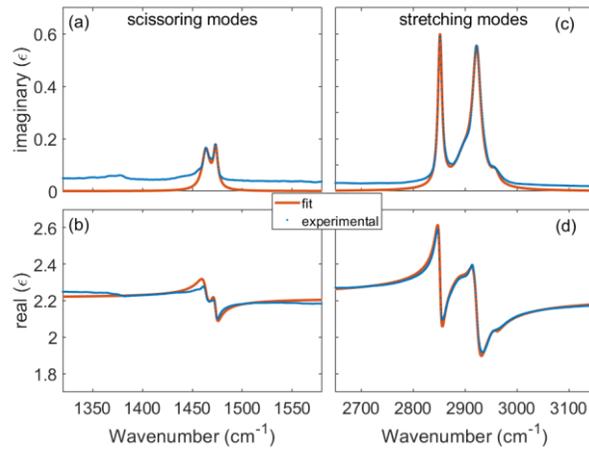

Figure 3: Fitting curves of the extracted dielectric function from the PE film using a multiple Lorentz oscillator permittivity model for both scissoring and stretching modes. On the upper panels are the imaginary parts, while the lower ones show the real part. Table in supplementary information provides the fitting values.

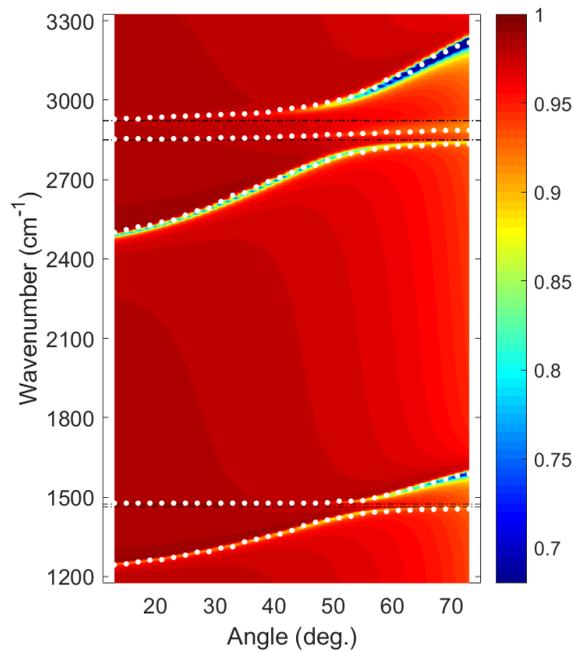

Figure 4: RCWA numerical simulation of the sample reflectance using the Lorentz oscillator permittivity fits. White dots are the experimental minima of reflectance extracted from the data, in excellent agreement with the simulations.

**Supplementary information:**

**Table 1:** parameters of the multiple Lorentz oscillator permittivity model

| Vibrational mode | $a_k$ (cm$^{-1}$) | $\omega_k$ (cm$^{-1}$) | $\gamma_k$ (cm$^{-1}$) |
|---|---|---|---|
| Scissoring 1 | 0.72 | 1464 | 9.3 |
| Scissoring 2 | 0.4 | 1473 | 5.3 |
| Stretching 1 | 2.7 | 2852 | 9.3 |
| Stretching 2 | 2.5 | 2899 | 43.4 |
| Stretching 3 | 4.8 | 2923 | 19.3 |
| Stretching 4 | 0.3 | 2959 | 16.7 |

**Figure S1:** Experimental angle-resolved reflectance of the 2.56μm PE film embedded in the double metal FP micro-cavity for different polarization condition of the impinging radiation.

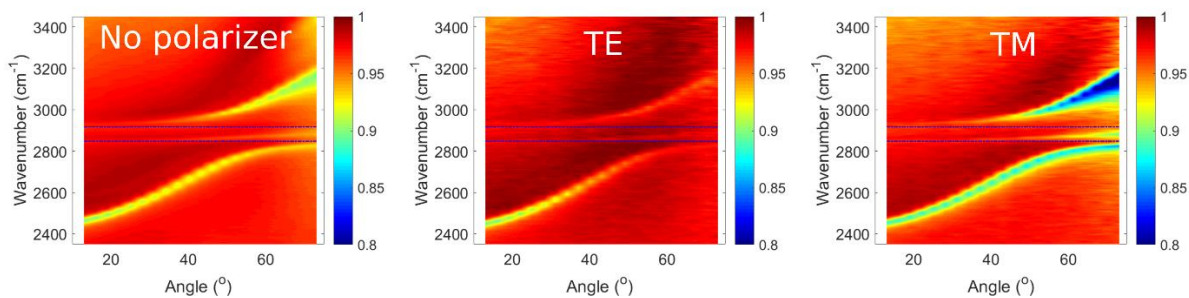

**Figure S2:** Reflectance of the strongly-coupled sample probed using a FTIR microscope with a 150 μm square focal spot. Probing at different position over a 1mm distance shows a 3$^{rd}$ FP mode resonant at different frequencies, which indicates a thickness disparity of about 160 nm.

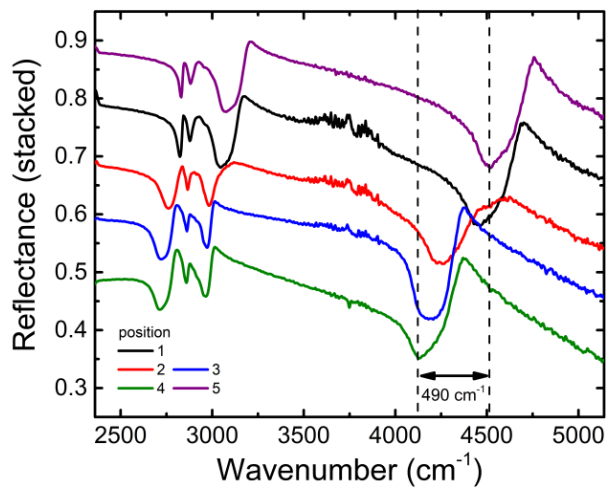